\documentclass[12pt,a4paper]{article}
\usepackage[]{latexsym,epsf}
\usepackage[xdvi]{graphicx}

\begin{document} 

\title{Non-exponential relaxation in diluted antiferromagnets}
\author{M.~Staats, U.~Nowak, K.~D.~Usadel\\ 
  Theoretische Tieftemperaturphysik\\ 
  Gerhard-Mercator-Universit\"{a}t-Duisburg\\
  47048 Duisburg/ Germany\\
  e-mail: michael@thp.uni-duisburg.de 
} 

\date{\today}
\maketitle

\begin{abstract}
  Diluted Ising antiferromagnets in a homogenous
  magnetic field have a disordered phase for sufficiently large
  values of the field and for low temperatures.  Here, the system is
  in a domain state with a broad size-distribution of fractal
  domains.  We study the relaxation dynamics of this domain state
  after removing the external field for two and three dimensions.
  Using Monte Carlo simulation techniques, we measure the decay of the
  remanent magnetization. Its temperature dependence can be understood
  as thermal activation.  All data can be described by a unique
  generalized power law for a wide range of temperatures in two and
  three dimensions. The question wether the exponent of the
  generalized power law is universal remains open.\\
\end{abstract}


\section{INTRODUCTION}
The dynamics of phase ordering of pure systems is quite well
understood.  Starting from a disordered state and quenching the system
 to temperatures well below the critical one, the growth of order is
characterized by a length $L(t)$ which scales with $t^{1/2}$ for any
spatial dimension $d$ (for a review see Bray, 1993).  However, the
effect of quenched disorder on the dynamics of phase ordering is less
well understood.  It is generally accepted that the dynamics of
disordered systems is driven by thermal activation following $t = \tau
\exp(\delta E/T)$, where $t$ is the time necessary for a physical
process that has to overcome energy barriers of height $\delta E$ and
$\tau$ a microscopic time scale of the system. Within a Monte Carlo
simulation, this time scale can be expected to be of the order of $1$
Monte Carlo Step (MCS).  This leads to a natural scaling variable of
$T \ln(t/\tau)$.  In theoretical works the focus is usually laid on
the characteristical length scale of the system such as the average
domain size (Fisher, 1993).  However, it is not always clear, in which
way physical quantities like the magnetization of a system can be
related to such length scales, since the complete distribution of
domain sizes might influence them (Nowak, Esser, and Usadel, 1996). In
experiments, usually the magnetization or the dynamic susceptibilities
are measured. In order to get results which are comparable to
experimental situations, we analyzed the decay of the remanent
magnetization of diluted Ising antiferromagnets, a model which is a
prototype for strongly disordered systems.

\section{MODEL AND SIMULATION}
We simulated two- and three-dimensional diluted antiferromagnets in an
external magnetic field. The Hamiltonian of the Ising system
($\sigma_i = \pm 1$) is given by
\begin{equation}
  {\cal H} = - J \sum_{<ij>} \varepsilon_i \varepsilon_j \sigma_i \sigma_j
  - B \sum_i \varepsilon_i \sigma_i,
\end{equation}
where the summation runs over the nearest neighbors on a square
($d=2$) and cubic ($d=3$) lattice, respectively. The interaction $J$
is set to $-1$. The dilution is given by the quenched variables
$\varepsilon_i$, which are $0$ with a probability $p$ and $1$
otherwise.

The DAFF is in the same universality class as the Random Field Ising
Model (RFIM) (Fishman and Aharony, 1979). Hence, for $D=2$ the system
is antiferromagnetically ordered only for $B=0$ below the critical
temperature $T_N$ (Aizenman and Wehr, 1989). In $D=3$ there is long
range antiferromagnetic order also for external fields $B>0$. In both
$D=2$ and $D=3$ the system is in a domain state for sufficiently large
$B$.  This is also true for $T=0$ as can be shown by exact ground
state calculations (Esser, Nowak, and Usadel, 1997). Here, the domains
of the DAFF are highly fractal, with a complex shape. Their size
distribution is broad, following a power-law with an exponential
cut-off which depends on the strength of the external field.
Experimental realizations of this model are e.\ g.\ 
$\mathrm{Fe_{1-p}Zn_pF_2}$ ($D=3$) and
$\textnormal{Rb}_2\textnormal{Co}_{1-p}\textnormal{Mg}_p\textnormal{F}_4$
($D=2$) (for are review of the experimental work, see Kleemann, 1993).

We investigate the dynamics of the DAFF with Monte Carlo techniques
using the heat-bath algorithm which simulates Glauber dynamics.  Spins
are updated in a random sequential way.  For simulation 
we averaged over 20 runs ($D=2$) with different realizations of the
dilution. The lattice size is $400 \times 400$ spins in 2D and $50
\times 50 \times 50$ in 3D with periodical boundary conditions.

\begin{figure}[bt]
  \begin{center}
    \includegraphics[angle = 90, height = 4cm]{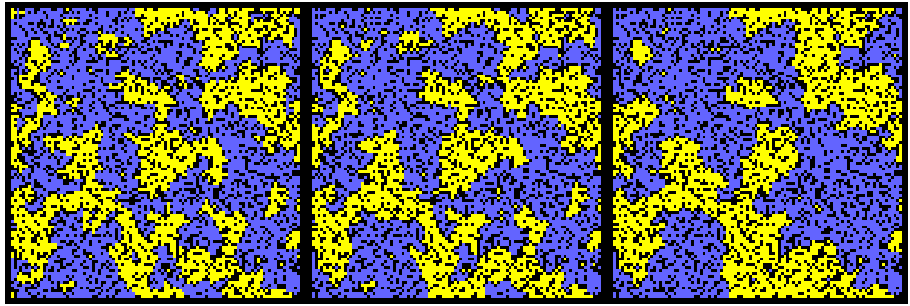}
  \end{center}
  \caption{Domain configurations during the relaxation after 1, 1023,
    and 262143 MCS (from left). The two antiferromagnetic phases are
    represented in yellow and blue, vacancies are black.}
  \label{f:relax}
\end{figure}

We prepare the system starting with a disordered spin configuration in
a region of the phase diagram where the equilibrium state of the
system is a domain state.  We performed a large number of Monte Carlo
updates until the homogeneous magnetization does not change
significantly any more.  Due to the slow dynamics, the system will not
have reached thermal equilibrium within our simulation time. However,
this is not important here. The only important aspect here is that the
system is in a typical domain state.  After this preparation, the
external field is switched off and the relaxation of the magnetization
is measured for different temperatures. The system will relax into its
long range ordered ground state with zero magnetization - which it
cannot reach within reasonable time scales due to the extremely slow
dynamics.  These non-exponential dynamics stems from the fact that the
relaxation of the system is due to domain wall movement.  The domain
walls are pinned to those sites with vacancies and due to this pinning
effect, energy barriers have to be overcome by thermal activation.
Hence, exponentially large time scales are involved in the dynamics of
the relaxation. Figure \ref{f:relax} shows three snap shots of the 2D
DAFF during the relaxation, at the beginning of the relaxation, for an
intermediate time, and at the end of our simulation time. 
The size of the system in Figure \ref{f:relax} is $100 \times 100$.
As one can see, and as was shown earlier quantitatively (Esser, Nowak,
and Usadel, 1997) the initial domain state of the system consists of
domains on all length scales.  During the relaxation, the smaller
domains vanish first (except for thermal fluctuation, which are also
visible since the pictures are snap-shots). At later times only large
domains still exist. Also, the domain walls are flattened. The
integral change of magnetization at some time $t$ is due to the sum of
all changes that happened during time - small changes for shorter time
and larger changes for longer times (Nowak, Esser, and Usadel, 1996).
The largest length scale involved in the dynamics can be expected to
be connected to the corresponding time scale $t$ by thermal activation
$t = \tau \exp (\delta E(L)/T)$.

\section{ANALYSIS OF SIMULATION DATA}
\begin{figure}[bt]
  \begin{center}
    \input{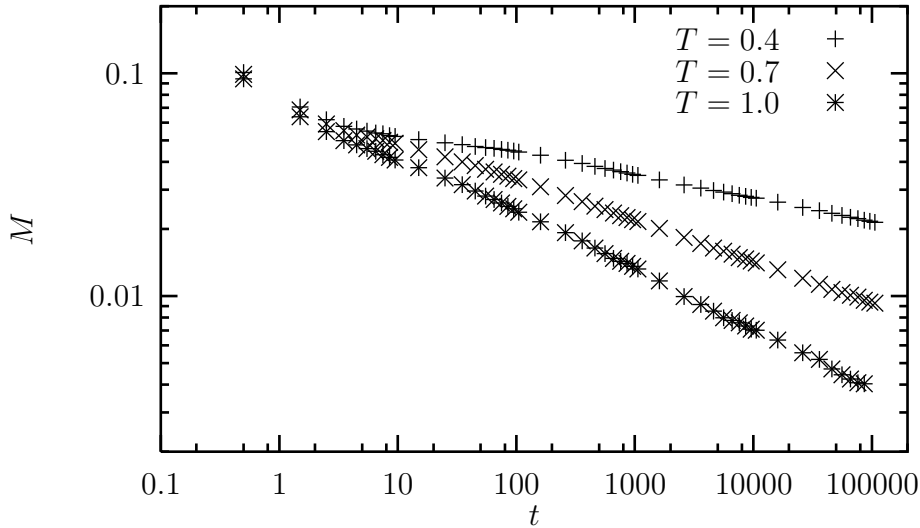}
  \end{center}
  \caption{Magnetisation versus time $t$ for the relaxation of a 2D
    DAFF for three different temperatures.}
  \label{f:unskal}
\end{figure}

\begin{figure}[bt]
  \begin{center}
    \input{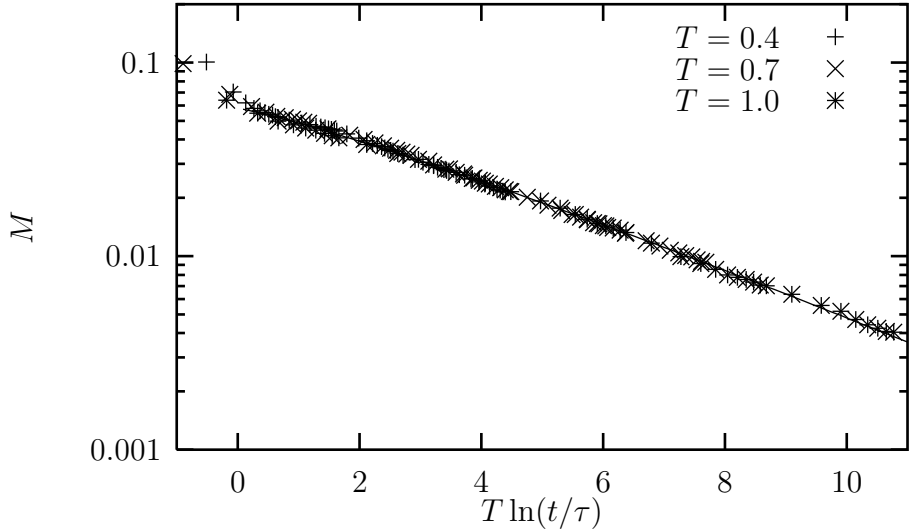}
  \end{center}
  \caption{Magnetisation versus scaling variable $T \ln (t/\tau)$ for
    the data of Figure \ref{f:unskal}.}
  \label{f:rem_2d}
\end{figure}

\begin{figure}[bt]
  \begin{center}
    \input{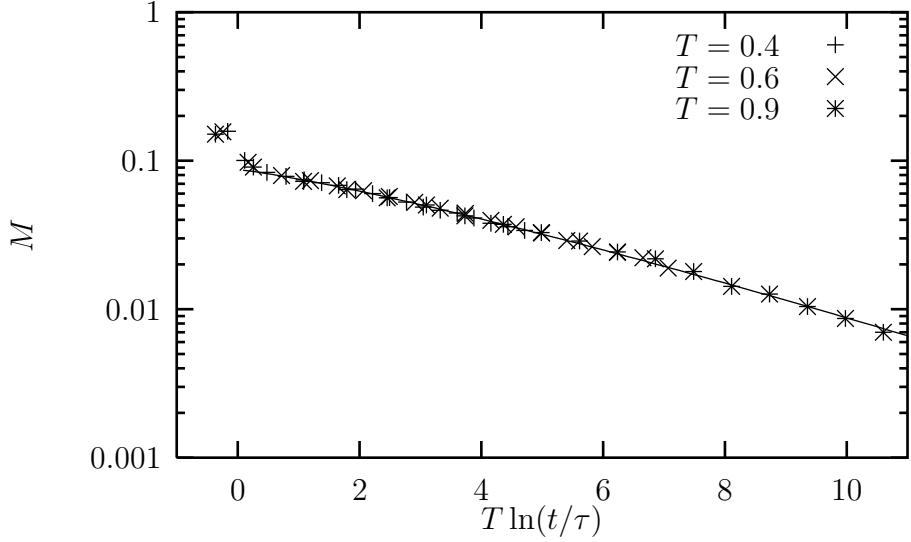}
  \end{center}
  \caption{Magnetisation versus scaling variable $T \ln (t/\tau)$ for
    the relaxation of a 3D DAFF.}
  \label{f:rem_3d}
\end{figure}

Figure \ref{f:unskal} shows the remanent magnetization versus time for
a two dimensional DAFF with an initial field of $B = 2.0$ and a
dilution of $p = 0.25$. Data for three different temperatures are
shown. As the semi-logarithmic plot suggests, the data follow roughly a
power-law with a temperature dependent exponent. However, 
a slight curvature is visible. Therefore, also other relaxation-laws
which are usually used to describe slow relaxation should be taken
into account.
To improve the accuracy of our analysis, we analyzed the data in two
steps. First, we consider for fixed external field $B$ and fixed
dilution $p$ all data for different temperatures. If, as we expect, $T
\ln(t/\tau)$ is the correct scaling variable all these data should
collapse on the same universal curve by a suitable choice of $\tau$.
Figure \ref{f:rem_2d} demonstrates that this works very well for the
2D system yielding $\tau = 1.8$.
The same analysis was done for the remanent magnetization of a 3D DAFF
with an initial field of $B = 2.0$ and a dilution of 50\%.
The corresponding scaling plot is shown in
Figure \ref{f:rem_3d} yielding $\tau = 1.5$.
Note, that we determined
the values of $\tau$ without any assumption about the decay law at
this stage of the analysis.

Having established that the decaying magnetization scales with $T
\ln(t/\tau)$ we now discuss its scaling function. It is not expected
that the magnetization decays as a simple power of the scaling length
as does presumably the characteristic domain size since there is no
obvious relation between the maximum relaxed domain size for a
certain time and the decay of magnetization during that time. A more
elaborate analysis (Nowak, Esser, and Usadel 1996) for the same
systems but for the dynamics of domain growth in a field starting with
long range order comes to the result that the change of 
magnetization in the limit of long times decays according to
\begin{equation}
\Delta M(t) = a \; e^{-b(T \ln(t/\tau))^y}.
\end{equation}
Therefore, we expect a similar behavior in our simulation. Apart from that,
experimental results (Han, Belanger, Kleemann, and Nowak, 1992) as
well as earlier simulations (Han and Belanger, 1992) for three
dimensional systems have also been fitted by this function.  With this
function a convincing fit of our time scaled data can be done as is
shown in Figures \ref{f:rem_2d} and \ref{f:rem_3d} by solid lines. The
resulting values for the parameters $\tau$ and the exponent $y$ are
summarized in Table \ref{results}. We did not find any significant
influence of the initial field $B$ on the relaxation.
\begin{table}[bt]
  \begin{center}
    \leavevmode
    \caption{Constants of the generalized power law for different
      dilutions in two and in three dimensions}
    \vspace*{5mm}
    
    \begin{tabular}{|c||c|c|c|c|c|c|}
      \hline
             &    \multicolumn{5}{c|}{2D}                    &    3D  \\ \hline
      $p$    & $0.15$ &  $0.2$ & $0.25$ &  $0.3$ & $0.35$         & $0.5 $ \\
      $y$    & $\approx 1$ & $\approx 1$ & $1.15$ & $1.3$ & $1.5$ & $1.2$ \\
      $\tau$ & $5$ & $3$ & $1.8$ & $1$ & $0.4$ & $1.2$ \\ \hline
    \end{tabular}
    \label{results}
  \end{center}
\end{table}

Through the introduction of the scaling variable $x = T \ln (t/\tau)$
we enhanced the range of data which can be analyzed. Hence, we can
differentiate the quality of several relaxation laws without any
fitting procedure. E. g., in terms of the scaling variable $x$, a pure
logarithmic law (see e.~g.~Nattermann and Vilfan, 1988) has the form
of a power law, $M(T,t) = M_0(p,B) x^{-\Psi}$.
\begin{figure}[bt]
  \begin{center}
    \input{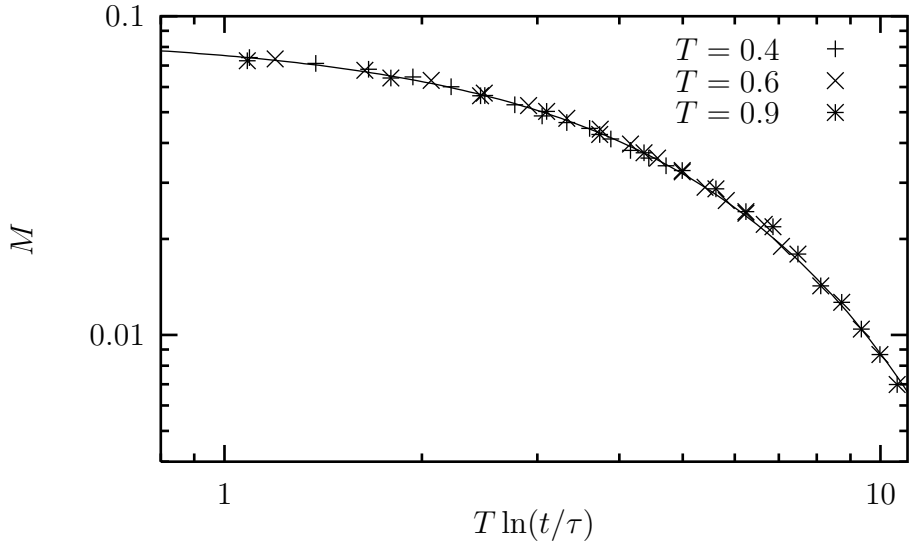}
  \end{center}
  \caption{Magnetisation versus scaling variable $T \ln (t/\tau)$ for
    the same data as in Figure \ref{f:rem_2d} on a logarithmic scale.}
  \label{f:natt}
\end{figure}
The corresponding plot in Figure \ref{f:natt} demonstrates, that this
description is not appropriate - otherwise we had a straight line in
the logarithmic plot.

\section{CONCLUSIONS}
We analyzed the non-exponential time dependence of the remanent
magnetization of diluted antiferromagnets. By introducing a scaling
variable $T \ln (t/\tau)$ we showed that the basic mechanism of the
relaxation is thermal activation. Also, we found that the relaxation
can very well be described by a generalized power-law for two as well
as for three dimensional systems. We did not find any significant
dependence of the relaxation on the initial magnetic field.

For two dimensional systems, we varied the dilution $p$ of the system.
The exponent $y$ of the generalized power law and the time constant
$\tau$ depend on the dilution of the system. However, we cannot
exclude the possibility that $y$ is a universal exponent and that the
$p$ dependence in our data stems from a crossover phenomenon due to the
percolation problem, because close to the percolation threshold $p_c$ a
crossover to a new kind of dynamics can be expected, since for
concentrations below $p_c$ no long-range order can be established in the
system.

{\bf Acknowledgments:} This work was in part supported by the
Deutsche Forschungsgemeinschaft through Sonderforschungsbereich 166. 

\section{References}
\begin{description}
\item Aizenman, M. and Wehr, J. (1989). Rounding of First-Order Phase
  Transition in Systems with Quenched Disorder.
  Phys. Rev. Lett. {\bf 62}, 2503.
\item Binder, K. and Heermann, D.~W. (1988). Monte Carlo Simulations
  in Statistical Physics. Springer-Verlag, Berlin.
\item Bray, A.~J. (1993). ``Domain growth and coarsening''. In: Phase
  Transitions and Relaxation in Systems with Competing Energy Scales,
  ed.~by Kluwer Academic Publishers, Dordrecht/NL, p.~405.
\item Bricmont, J. and Kupiainen, A. (1987). Lower critical dimension for
  the Random-Field Ising model. Phys.~Rev.~Lett. {\bf 59}, 1829.
\item Esser, J., Nowak, U., and Usadel, K.~D. (1997). Exact Ground State
  Properties of Disordered Ising Systems. Phys.~Rev.~B {\bf 55}, 5866.
\item Fisher, D.~S. (1993). ``Low temperature phases, ordering and dynamics in
  random media''. In: Phase Transitions and Relaxation in Systems with
  Competing Energy Scales, ed.~by Kluwer Academic Publishers,
  Dordrecht/NL, p.~1.
\item Fishman, S. and Aharony, A. (1979). Random field effects in disordered
  anisotropic antiferromagnets. J.~Phys. {\bf  C12}, L729.
\item Han, S.~J. and Belanger, D.~P. (1992). Relaxation of the remanent
  magnetization of dilute anisotropic antiferromagnets. Phys.~Rev.~B
  {\bf 46},  2926.
\item Han, S.~J., Belanger, D.~P., Kleemann, W., and Nowak, U. (1992).
  Relaxation of the excess magnetization of random-field-induced
  metastable domains in $\mbox{Fe}_{0.47}\mbox{Zn}_{0.53}\mbox{F}_2$.
  Phys. Rev. B {\bf 45}, 9728.
\item Imbrie, J.~Z. (1984). Lower critical dimension of the Random-Field
  Ising model. Phys. Rev. Lett. {\bf 53}, 1747.
\item Kleemann, W. (1993). Random-Field induced antiferromagnetic,
  ferromagnetic and structural domain states. Int.~J.~Mod.~Phys.~B {\bf 7},
  2469.
\item Nattermann, T. and Vilfan, I. (1988). Anomalous Relaxation in the
  Random-Field Ising Model and Related Systems. Phys. Rev. Lett. {\bf
  61}, 223.
\item Nowak, U., Esser,~J., and Usadel, K.~D. (1996). Dynamics of
  domains in diluted antiferromagnets. Physica A {\bf 232}, 40.  
\end{description}
\end{document}